\documentstyle[preprint,prl,aps,epsfig]{revtex}

\begin{document}
\draft
\begin{titlepage}
\preprint{\vbox{\hbox{UM-P-98/45} \hbox{UDHEP-09-98} 
\hbox{TMUP-HEL-9812} \hbox{September 1998}}}
\title{\large \bf Atmospheric Neutrino Tests of Neutrino 
Oscillation Mechanisms}
\author{\bf R. Foot$^{(a)}$, C. N. Leung$^{(b)}$, 
and O. Yasuda$^{(c)}$} 
\address{(a) School of Physics, Research Centre for High 
Energy Physics\\
The University of Melbourne, Parkville 3052 Australia \\}
\address{(b) Department of Physics and Astronomy, 
University of Delaware\\
Newark, DE 19716, U.S.A. \\}
\address{(c) Department of Physics, Tokyo Metropolitan 
University\\ 
1-1 Minami-Osawa Hachioji, Tokyo 192-0397, Japan\\} 
\maketitle
\bigskip
\begin{abstract} 
 
Recent Super-Kamiokande data on the atmospheric neutrino 
anomaly are used to test various mechanisms for neutrino 
oscillations.  It is found that the current atmospheric 
neutrino data alone cannot rule out any particular 
mechanism.  Future long-baseline experiments should play
an important role in identifying the underlying neutrino 
oscillation mechanism.  

\end{abstract}
\end{titlepage}

\newpage
The atmospheric neutrino anomaly\cite{atm} has been 
confirmed by the Super-Kamiokande experiment\cite{SK}.  
The observed up-down asymmetries of the detected muons
indicate that, with the three known neutrino flavours, 
the most likely solution to this anomaly is $\nu_\mu 
\to \nu_\tau$ oscillations, although significant 
mixing with $\nu_e$ cannot yet be excluded\cite{fvy3}.  

The phenomenon of neutrino flavour oscillations was first 
proposed as a consequence of nondegenerate neutrino masses
\cite{Ponte}.  Although this mass mixing of weak eigenstates 
is the most likely mechanism for the observed atmospheric 
neutrino oscillations, other possible mechanisms have been 
proposed and, as stressed in Ref.\cite{H}, it is important 
to let experiments rather than our theoretical prejudice 
determine which is the correct mechanism. It is in this 
spirit that we undertake the present study.  

The alternative neutrino oscillation mechanisms considered 
below share a common feature: each requires the existence 
of an interaction (other than the neutrino mass terms) that 
can mix neutrino flavours.  In this paper we shall focus on 
such interactions which are mediated by a scalar ($J = 0$), 
a vector ($J = 1$), or a tensor ($J = 2$) field.  Assuming 
a two-neutrino mixing scheme, the $\nu_\mu \rightarrow 
\nu_\tau$ transition probability for each of these 
possibilities may be parametrized as \cite{Y}
\begin{equation}
 P(\nu_\mu \rightarrow \nu_\tau) = \sin^2 2 \theta_J
 \sin^2(E_\nu^{J-1} L \delta_J),
\label{p}
\end{equation}
where $E_\nu$ is the neutrino's energy, $L$ is the neutrino's 
path length (i.e., the distance from where the neutrino is 
produced to the detector), $\delta_J$ is a parameter specific 
to the neutrino oscillation mechanism, and $\theta_J$ is the 
corresponding mixing angle.  The subscript $J$ in $\theta_J$ 
and $\delta_J$ is simply a label for the different mechanisms, 
and does not imply that the values of these parameters depend 
on the value of $J$.  Note the distinct energy dependence of 
the oscillation probability for each value of $J$.  This is 
the key for determining which neutrino oscillation mechanism 
is at work.  

Proposed mechanisms for the $J = 0$ case include the mass 
mixing mechanism\cite{Ponte} for which the parameter 
$\theta_0$ in Eq.(\ref{p}) represents the mixing angle 
between the neutrino flavour eigenstates and the mass 
eigenstates, and the parameter $\delta_0$ is given by 
\begin{equation}
 \delta_0({\rm mass}) = \frac{\Delta m^2}{4} \equiv 
 \frac{m_2^2 - m_1^2}{4}_,
\label{deltam}
\end{equation}
where $m_i$ are the masses of the neutrino mass eigenstates.  
Another possibility for the $J = 0$ case is neutrino 
oscillations induced by nonuniversal couplings of the 
neutrinos to a massless string dilaton for which 
$\delta_0$ is given by\cite{sd} 
\begin{equation}
 \delta_0({\rm dilaton}) = \frac{1}{4} \left[\Delta m^2 - 
 2 \phi (\alpha_2 m_2^2 - \alpha_1 m_1^2) \right]. 
\label{delta0}
\end{equation}
Here $\alpha_i$ are the coupling strengths of different 
neutrino gravitational eigenstates to the dilaton field 
and it has been assumed that the neutrino mass eigenstates 
and the gravitational eigenstates are the same.  The 
angle $\theta_0$ for this case is therefore the mixing 
angle between the flavour eigenstates and the gravitational 
eigenstates.  Note that the nonuniversal neutrino-dilaton 
couplings constitute a violation of the principle of 
equivalence.  The parameter $\phi$ in Eq.(\ref{delta0}) 
denotes the local Newtonian gravitational potential.  It is 
assumed to be constant over the neutrino path length since 
the dominant contribution to $\phi$ appears to come from 
the great attractor\cite{HLP} which has been estimated to 
be\cite{Kenyon}: $|\phi| \sim 3 \times 10^{-5}$.  With the 
constant $\phi$ approximation, the energy dependence in the 
oscillation probability is the same for this mechanism as 
it is for the mass mixing mechanism.  Neutrino oscillation 
experiments alone will not be able to distinguish these 
two possibilities.  For the purposes of this paper, we shall 
refer to them collectively as the scalar mechanism.  Note, 
however, that the mass mixing mechanism requires the neutrino 
masses to be nondegenerate whereas the dilaton induced 
oscillations can take place even if the masses are degenerate.  
We also observe that, if the neutrino-dilaton couplings are 
universal, i.e., if $\alpha_1 = \alpha_2$, and if $\delta_0
({\rm mass}) \neq 0$, $\delta_0({\rm dilaton})$ and 
$\delta_0({\rm mass})$ differ only by a multiplicative 
constant.

As an example of the $J = 2$ case, consider neutrino 
oscillations induced by the equivalence principle violation 
that results from nonuniversal couplings of neutrinos to 
gravity\cite{VEP}.  In this case $\delta_2$ is given 
by\cite{HLP} 
\begin{equation}
 \delta_2({\rm VEP}) = |\phi| \Delta \gamma, 
\label{delta2}
\end{equation}
where $\Delta \gamma$ measures the degree of violation of 
the equivalence principle and, as in Eq.(\ref{delta0}), 
$\phi$ denotes the essentially constant local gravitational 
potential.  With a constant $\phi$, this mechanism is 
phenomenologically identical\cite{GHKLP} to the velocity 
oscillation mechanism that arises from a breakdown of Lorentz 
invariance\cite{CG} for which $\delta_2$ assumes the form
\begin{equation}
 \delta_2({\rm velocity}) = \frac{\Delta v}{2}_,
\label{deltav}
\end{equation}
where $\Delta v = v_2 - v_1$ is the difference between the 
velocities of two neutrino velocity eigenstates.  We shall 
refer to these two mechanisms collectively as the tensor 
mechanism.  In the former case, the angle $\theta_2$ 
corresponds to the mixing angle between the flavour 
eigenstates and the gravitational eigenstates, whereas in 
the latter case $\theta_2$ corresponds to the mixing angle 
between the flavour eigenstates and the velocity eigenstates.  
It should be emphasized that the tensor mechanism can lead 
to neutrino oscillations even if neutrinos are massless (or 
degenerate).  These mechanisms were proposed not so much as 
a competing mechanism for the mass mixing mechanism, but to 
point out that neutrino oscillation experiments could be 
used as high-precision tests of the symmetry principles 
fundamental to the theories of general and special relativity.  

We include the vector mechanism (the $J = 1$ case in 
Eq.(\ref{p})) in our phenomenological study.  Although 
there are models which can yield energy-independent neutrino 
oscillations in matter\cite{FCNC,Gasperini,FY}, we are not 
aware of any such model that can explain the atmospheric 
neutrino anomaly.  The aim of our study is not to test any 
specific model, but rather to check if an energy-independent 
oscillation mechanism is compatible with the atmospheric 
neutrino data. 

To test the three classes of neutrino oscillation mechanisms, 
we compare the measured values of  
\begin{equation}
 R \equiv \frac{(N_\mu/N_e)|_{\rm data}}
 {(N_\mu/N_e)|_{\rm MC}}
\label{R}
\end{equation}
and the up-down asymmetry parameters 
\begin{equation}
 Y_\alpha^\eta \equiv \frac{(N_\alpha^{- \eta}/N_\alpha^{+ \eta})
 |_{\rm data}}{(N_\alpha^{- \eta}/N_\alpha^{+ \eta})|_{\rm MC}}_,
 ~~~~~~\alpha = e, \mu
\label{Y}
\end{equation}
with the corresponding predictions of these mechanisms, assuming 
$\nu_\mu \rightarrow \nu_\tau$  transitions.  Here $N_e$ and 
$N_\mu$ are the number of $e$-like and $\mu$-like events, 
respectively.  $N_\alpha^{-\eta}$ and $N_\alpha^{+\eta}$ are the 
number of $\alpha$-like events produced in the detector with 
zenith angle $\cos \Theta < - \eta$ and $\cos \Theta > \eta$, 
respectively.  Note that $\Theta$ is defined to be negative for 
upward going directions and we have chosen $\eta = 0.2$ in our 
analysis.  The calculational method is identical to that described 
in \cite{fvy} and will not be reproduced here for environmental 
reasons.  

Our results for the $R$'s and $Y$'s are displayed in Figures 
1 and 2 together with the Super-Kamiokande results \cite{SK},
\begin{eqnarray}
R ({\rm sub-GeV}) & = & 0.63 \pm 0.03 \pm 0.05, \nonumber\\
R ({\rm multi-GeV}) & = & 0.65 \pm 0.05 \pm 0.08, \nonumber\\
Y^{0.2}_{\mu} ({\rm sub-GeV}) & = & 0.79 \pm 0.05, \nonumber\\
Y^{0.2}_{\mu} ({\rm multi-GeV}) & = & 0.56 \pm 0.06. 
\end{eqnarray}
The experimental results we use correspond to 535 live 
days of running.  For completeness we mention that the 
Super-Kamiokande results for the $e$-like up-down asymmetries 
are
\begin{eqnarray}
Y^{0.2}_{e} ({\rm sub-GeV}) & = & 1.14 \pm 0.07, \nonumber\\
Y^{0.2}_{e} ({\rm multi-GeV}) & = & 0.92 \pm 0.12.
\end{eqnarray}
Finally, note that only statistical errors are given for the 
up-down asymmetries since they should be much larger than 
possible systematic errors at the moment.  We have not shown 
any figure for $Y_e$ - it is expected to be about 1 
because $\nu_\mu \to \nu_\tau$ oscillations have almost no 
effect on the expected number of electron events.

From Figures 1 and 2 we see that all three mechanisms can fit 
the data for a range of the parameter $\delta_J$.  The combined 
$\chi^2$ fit to $Y_\mu, Y_e$ and $R$ yield the allowed regions 
for the neutrino oscillation parameters shown in Figure 3.  
The $\chi^2$ function is defined to be 
\begin{equation}
\chi^2_{\rm atm} \equiv \chi^2(R)+\chi^2(Y),
\label{chi2}
\end{equation}
where 
\begin{equation}
\chi^2(R)\equiv\sum_E \left[\left({R^{SK} - R^{th} \over
\delta R^{SK}}\right)^2\right]
\end{equation}
and
\begin{equation}
\chi^2(Y)\equiv\sum_E \left[\left({Y^{SK}_{\mu} - Y^{th}_{\mu}
\over \delta Y^{SK}_{\mu}}\right)^2
+ \left({Y^{SK}_{e} - Y^{th}_{e} \over \delta Y^{SK}_{e}}
\right)^2 \right]_.
\end{equation}
The sum is over the sub-GeV and multi-GeV data samples.  The 
measured Super-Kamiokande values and errors are denoted by 
the superscript ``SK'' and the theoretical predictions for the 
corresponding quantities are labelled by the superscript ``th''.  
The minimal $\chi^2$ for the scalar, vector, and tensor 
mechanisms is 5.4, 5.7, and 10.4, respectively, for 4 degrees 
of freedom.  All three mechanisms are consistent with the 
data, but the tensor mechanism has the worst $\chi^2$.  

For check we have performed a second $\chi^2$ analysis 
using the SK $\chi^2$ function, 
\begin{equation}
\chi^2={\alpha^2 \over \sigma_\alpha^2}
+{\beta^2 \over \sigma_\beta^2}
+\chi_{\rm sub-GeV}^2+\chi_{\rm multi-GeV}^2,
\label{chi2sk}
\end{equation}
where
\begin{equation}
\displaystyle\chi_{\rm sub-GeV}^2 =
\sum_{\alpha=e,\mu}\sum_{a=1}^5
{\left( y_a^\alpha-n_a^\alpha\right)^2 \over n_a^\alpha}
\end{equation}
and similarly for $\displaystyle\chi_{\rm multi-GeV}^2$.  
Here $\alpha$ and $\beta$ are the absolute and relative 
normalizations of the neutrino flux, $y_a^\alpha$ are 
theoretical predictions for each zenith angle bin,
$n_a^\alpha$ are the experimental data, and we assume 
the uncertainties are given by $\sigma_\alpha=\infty$ and 
$\sigma_\beta=0.12$ (we assumed the larger value of
$\sigma_\beta$ given in the third reference of \cite{SK}).
It is understood that $y_a^\mu$ ($\mu$-like events) are 
multiplied by $(1+\alpha)(1+\beta/2)$ while $y_a^e$ 
(e-like events) are multiplied by $(1+\alpha) (1-\beta/2)$, 
and $\chi^2$ is optimized with respect to $\alpha$ and $\beta$.  
With this alternative $\chi^2$, we also found that all three 
cases ($J=0,1,2$) gave an acceptable fit to the data.

The basic reason that the Super-Kamiokande atmospheric 
neutrino data cannot distinguish these rather different 
oscillation probabilities is geometrical.  Neutrinos
from above typically travel quite short distances 
($\lesssim 50 \ {\rm km}$) whereas neutrinos
going up through the Earth travel quite long distances 
($\gtrsim 5000 \ {\rm km}$).  The atmospheric 
neutrino data can be explained by assuming that neutrinos 
from above do not have time to oscillate while neutrinos 
travelling through the Earth experience averaged 
oscillations.  In fact the data suggest that this occurs
for both the sub-GeV and multi-GeV energy range (which
is, very roughly, $0.3~{\rm GeV}\lesssim E_\nu 
\lesssim 6~{\rm GeV}$).  Thus, while atmospheric 
neutrino experiments can sensitively test for oscillations  
through the zenith angle dependence (or up-down asymmetry) 
they are really only sensitive to averaged oscillations 
and are consequently not sensitive to the oscillation 
mechanism.  The explicit energy dependence of the 
oscillation probability is only important in the region 
where the oscillations are significant and are {\it not} 
averaged.  The proposed long-baseline experiments should 
be much more sensitive to the oscillation mechanism.
The reason is that the neutrino path length is fixed and
at a moderate distance (typically about 250 - 600 km).  
Provided that not all of the oscillations are averaged, 
and provided that at least some of the neutrinos have 
time to oscillate, then a comparison of the detected energy 
distribution with the expected energy distribution should 
distinguish the oscillation mechanisms.

In summary, we have examined three distinct oscillation
mechanisms ($J=0,1,2$ in Eq.(\ref{p})) and compared these
with the current Super-Kamiokande atmospheric neutrino 
data.  We have found that the current data are quite insensitive 
to the oscillation mechanism responsible for the atmospheric 
neutrino anomaly.  To ultimately determine the underlying 
neutrino oscillation mechanism, it is important to do a more 
controlled experiment.  This should be possible in the near 
future with the advent of long-baseline experiments.

\newpage
\centerline{\bf Acknowledgement}
\bigskip
C.N.L. was supported in part by the U.S. Department of Energy 
under grant DE-FG02-84ER40163.  O. Y. was supported in part by a
Grant-in-Aid for Scientific Research of the Ministry of Education,
Science and Culture, \#09045036, \#10140221 and \#10640280.  R. F. 
is an Australian Research Fellow.

\newpage

\begin{center}
{\bf FIGURE CAPTIONS}
\end{center}
\vskip 0.8cm
\noindent
{\bf Figure 1:}  $R$ as a function of $\delta_J$ (see 
Eq.(\ref{p})) for the tensor (bold solid line), vector 
(solid line), and scalar (dashed line) mechanisms.  Maximal 
mixing has been assumed in each case.  The dimensionless 
variable X is equal to $\delta_2$ GeV-km, $\delta_1$ km, and 
$\frac{\delta_0}{1.27} ~{\rm GeV}^{-1}$-km, respectively for 
the tensor, vector, and scalar mechanisms (the choice of X 
for the mass mixing mechanism corresponds to the familiar 
$\Delta m^2/{\rm eV}^2$).  The horizontal dashed lines 
correspond to Super-Kamiokande's measured $R$ value 
$\pm$ 1$\sigma$ error.  Figure 1a is for the sub-GeV 
neutrinos and Figure 1b is for the multi-Gev neutrinos.

\vskip 0.5cm
\noindent
{\bf Figure 2:} Same as Figure 1 except $Y_\mu$ (see 
text) is plotted instead of $R$.

\vskip 0.5cm
\noindent
{\bf Figure 3:} The allowed region for the mixing parameter 
$\theta_J$ and the oscillation parameter $\delta_J$ obtained 
with the $\chi^2$ function defined in Eq.(\ref{chi2}).  
Figure 3a, 3b, and 3c are for the scalar, vector, and tensor 
mechanisms, respectively.

\newpage
\pagestyle{empty}
\epsfig{file=fig1a.eps,width=15cm}
\newpage
\epsfig{file=fig1b.eps,width=15cm}
\newpage
\epsfig{file=fig2a.eps,width=15cm}

\newpage
\epsfig{file=fig2b.eps,width=15cm}
\newpage
\epsfig{file=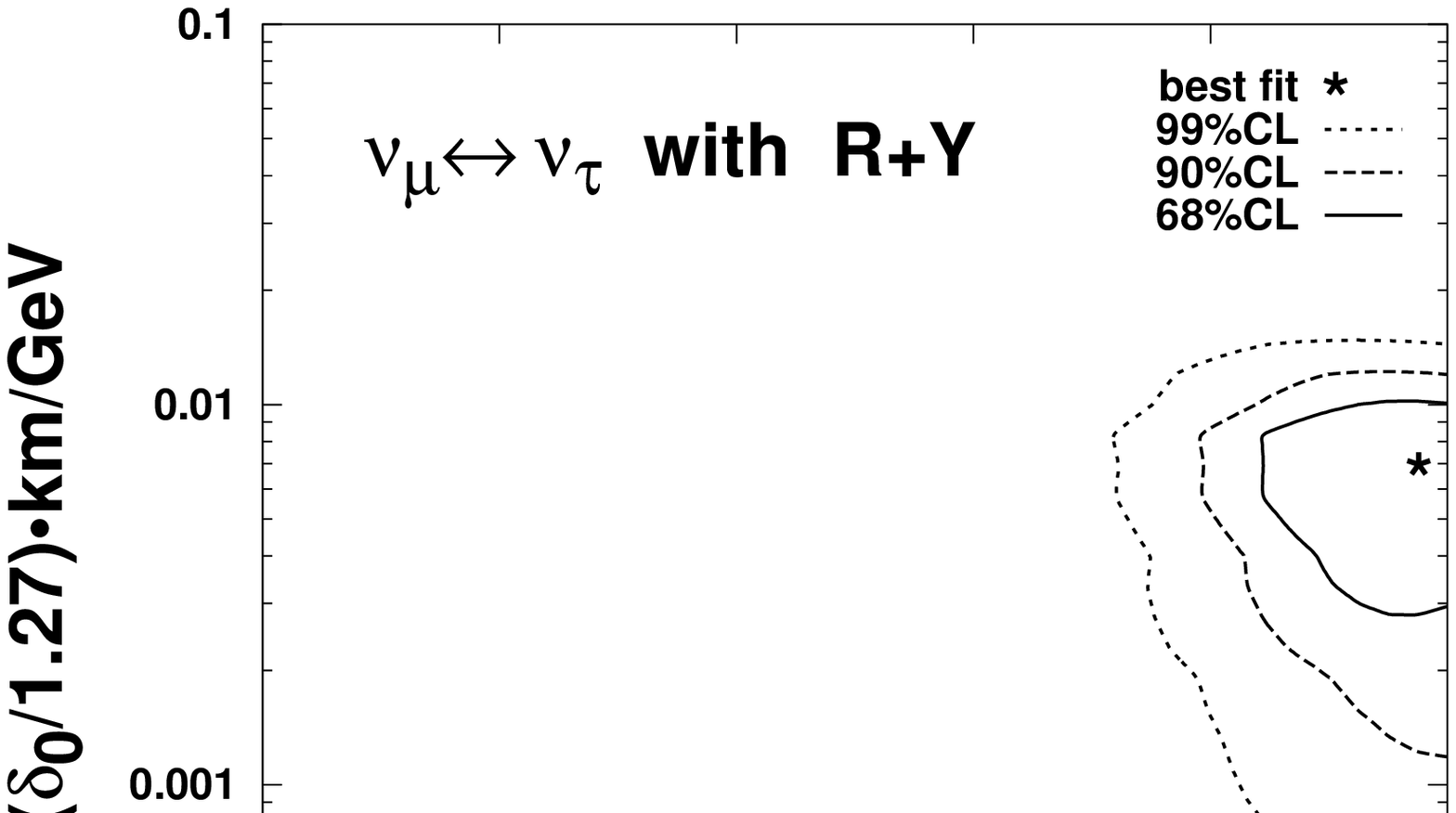,width=15cm}
\newpage
\epsfig{file=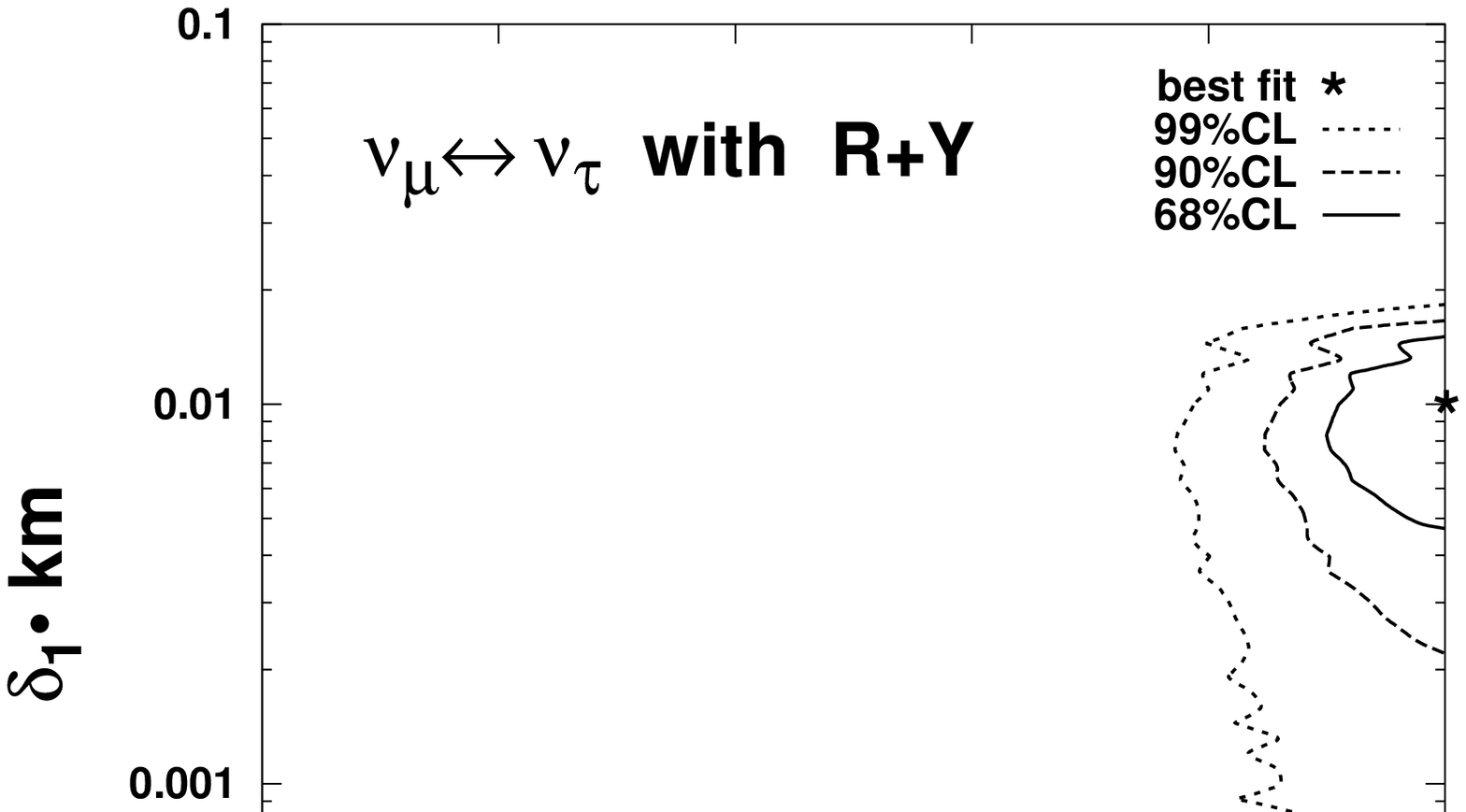,width=15cm}
\newpage
\epsfig{file=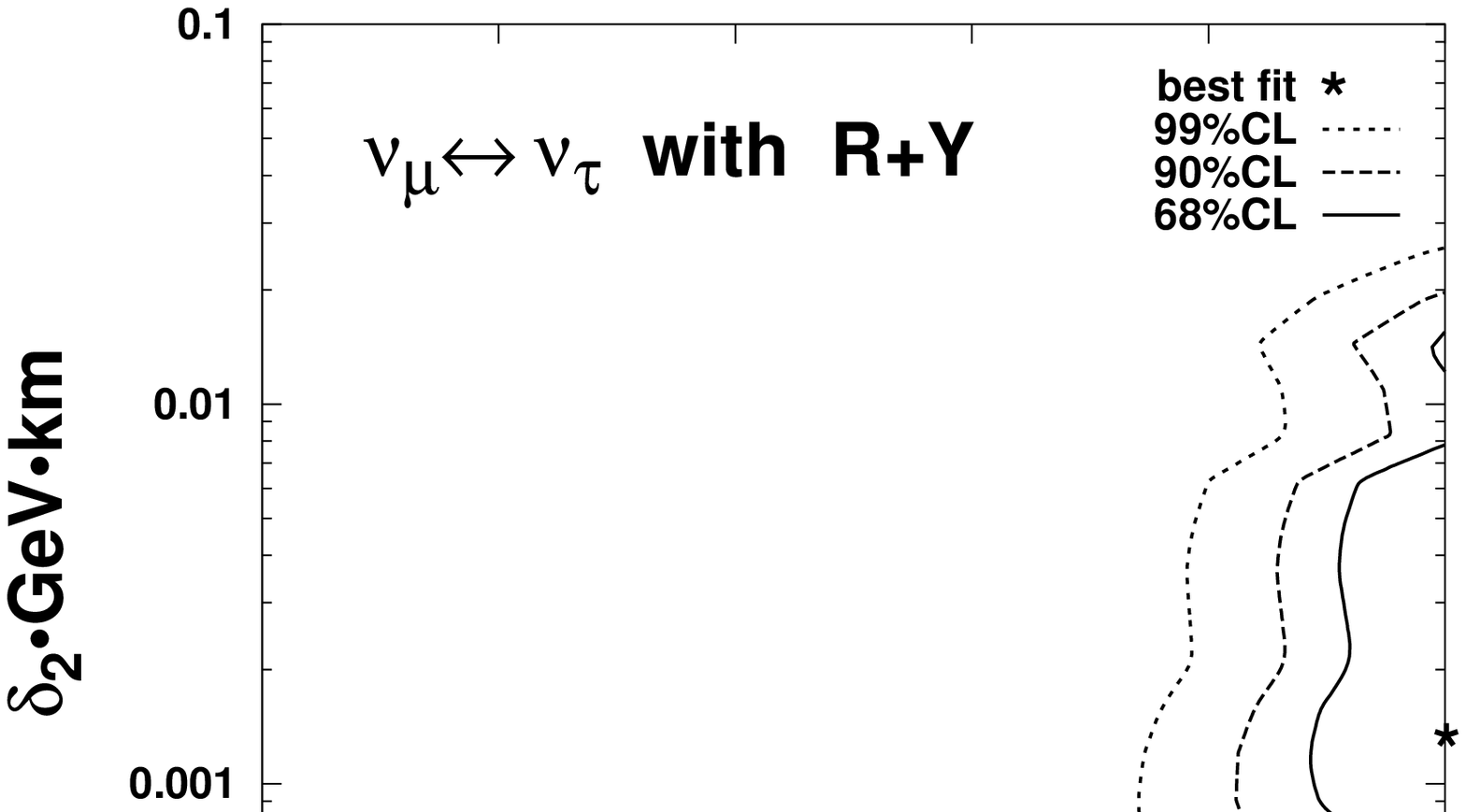,width=15cm}

\end{document}